# ADEPT: Architecture-Driven Energy-Efficient CNN Fine-Tuning on PIM Accelerators

Pratyush Dhingra, *Graduate Student Member, IEEE,* Vibhanshu Sharma, *Graduate Student Member, IEEE,* Janardhan Rao Doppa, *Senior Member, IEEE, and* Partha Pratim Pande, *Fellow, IEEE*

***Abstract*—Processing-in-memory-based (PIM) architectures have emerged as a promising solution for accelerating Convolutional Neural Network (CNN) workloads at the edge. Fine-tuning pre-trained CNNs is a common requirement to enhance the model's predictive accuracy after deployment. However, the fine-tuning process is computational and memory-intensive, generating a significant amount of intermediate activations. This leads to frequent off-chip memory access, affecting the overall efficiency of the PIM accelerator. Existing fine-tuning strategies are agnostic to the underlying hardware, as they treat all layers equally. In this paper, we propose a hardware-aware framework called ADEPT to accelerate CNN fine-tuning on PIM architectures. Unlike prior fine-tuning methods, ADEPT adaptively trains the model considering both the training overhead and layer sensitivity. Specifically, ADEPT introduces a novel metric that quantifies the trade-off between a block's gradient-based sensitivity and its hardware architecture-specific Energy-Delay Product (EDP), producing platform-dependent fine-tuning configurations. Overall, ADEPT helps reduce the total trainable parameters and the off-chip data access during fine-tuning, while incurring minimal loss in predictive accuracy compared to full-parameter fine-tuning.**

## I. INTRODUCTION

Convolutional Neural Networks (CNNs) are increasingly being deployed on edge devices across a variety of domains [1], [2]. While Vision Transformers dominate server-class workloads, CNNs are preferred for resource-constrained edge devices due to their superior inductive biases and lower parameter count [3]. However, edge devices generate a vast amount of data whose distribution can change over time compared to the training data [4]. The models pre-trained using a specific data distribution may need to be updated to achieve high predictive accuracy [5]. Moreover, there is an increasing demand for fine-tuning pre-trained CNNs directly at the edge due to concerns over data privacy [6].

Critically, on-device fine-tuning is not a one-time event. In federated learning deployments, each edge device performs multiple rounds of local fine-tuning per communication round [7]. Similarly, continual learning scenarios necessitate periodic model adaptation as the data distribution evolves over time, such as autonomous vehicles encountering new driving conditions or medical devices adapting to patient-specific physiology [8]. In these settings, fine-tuning constitutes a recurring and high energy consuming workload. Moreover, unlike inferencing, fine-tuning is compute-heavy and memory-intensive. Hence, enhancing the energy efficiency of on-device fine-tuning is essential for practical edge deployment.

Processing-in-memory (PIM)-enabled architectures have enabled high-performance and energy-efficient CNN inferencing [9], [10], [11]. Despite their advantages, PIM architectures face significant challenges in enabling the back-propagation computation required for model fine-tuning [11]. The back-propagation phase requires activations generated during forward-propagation for weight updates. The on-chip buffers (cache) are mostly insufficient to store the large number of intermediate activations necessary for weight updates. For example, during mini-batch training with a batch size of 16, MobileNet-V2 generates over 1 GB in intermediate activations [12]. This large activation size results in off-chip access, which leads to significant performance and energy overheads [13]. Hence, we need to explore novel fine-tuning techniques.

A common approach for efficient fine-tuning is to train the last few layers of a pre-trained model, preserving learned features while adapting to the new task [14], [15], [16]. However, these static heuristics are fundamentally agnostic to the target data distribution, leading to poor generalization under distribution shifts [17]. Specifically, the type of target data distribution compared to the pre-training data influences which layers are more effective to train, as we show later. Additionally, Parameter-Efficient Fine-Tuning (PEFT) methods, such as *Low-Rank Adaptation* (LoRA), have also emerged as an efficient fine-tuning technique [5]. Despite its success, the development of PEFT methods has predominantly centered on transformers or LLMs where LoRA is utilized effectively on the large, dense linear layers (e.g., $1024 \times 1024$). Here, a low-rank decomposition (e.g., rank of 8) of pre-trained weights yields a substantial reduction in trainable

This work was supported by the US National Science Foundation (NSF) grant CSR-2308530, the Army Research Office under Grant ARO-W911NF-24-1-0240 and the U.S. Department of Energy, Office of Science, Advanced Scientific Computing Research program under project 84245—Democratization of Co-design for Energy-Efficient Heterogeneous Computing (DeCoDe) at Pacific Northwest National Laboratory (PNNL). PNNL is a multi-program national laboratory operated for the U.S. Department of Energy (DOE) by Battelle Memorial Institute under Contract No. DE-AC05-76RL01830. *(Corresponding author: Pratyush Dhingra.)*
The authors are with the Department of Electrical Engineering and Computer Science, Washington State University, Pullman, WA 99163 USA (e-mail: pratyush.dhingra@wsu.edu; vibhanshu.sharma@wsu.edu; jana.doppa@wsu. edu; pande@wsu.edu).

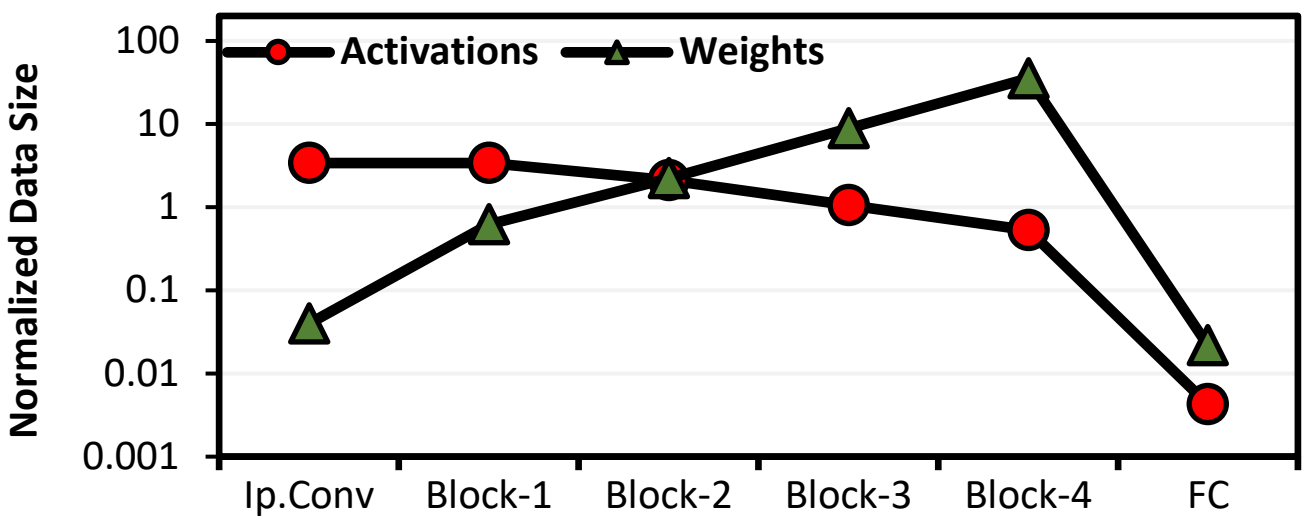

**Fig. 1:** The intermediate size of output activations and weights for input conv (Ip.conv), each conv block and the FC layer in ResNet-18.

parameters. In contrast, convolutional filters in CNNs typically have small dimensions, e.g., $3 \times 3$, thus making the standard application of LoRA unsuitable.

As highlighted, the critical bottleneck during fine-tuning on PIM architectures arises from the off-chip data access. CNNs exhibit a distinct structural pattern, unlike LLMs or transformers. CNNs process high-dimensional image data by progressively transforming input features into abstract representations, characterized by the reduction of spatial dimensions (width and height) [1]. Additionally, CNNs increase the depth or the number of channels across successive layers to capture richer and more complex hierarchical features [1], [18]. Fig. 1 shows the variation in size of forward activations and the weight parameters across the input convolutional (conv) layer, conv blocks, and fully-connected (FC) layer in ResNet-18 as an example. During the back-propagation phase, computing weight gradients necessitates fetching the forward activations of the preceding layer. Since early CNN layers possess massive spatial dimensions, retrieving these high-resolution activations from off-chip DRAM imposes substantial energy requirements. Conversely, the later convolutional layers feature expanded channel dimensions (Fig. 1), shifting the off-chip memory bottleneck from activation fetching to the gradients required for weight updates. Consequently, we need to capture the properties of each layer to effectively accelerate CNN fine-tuning.

In this work, we propose a hardware-aware energy-efficient CNN fine-tuning framework for PIM architectures, referred to as ADEPT. Unlike prior fine-tuning methods, ADEPT is the first to co-optimize layer sensitivity with actual hardware Energy-Delay Product (EDP) cost. ADEPT consists of two novel, synergistic components. First, ADEPT determines an optimized subset of CNN blocks to fine-tune on a per-batch basis. This selection is guided by a scoring metric that determines the relative sensitivity of each block against the overhead of training it, producing architecture-specific fine-tuning configurations. Second, we introduce a tailored Channel-wise LoRA (DCLoRA) that applies low-rank adaptation across the channel dimensions instead of filters within a conv layer. Unlike conventional LoRA, the proposed adaptation effectively reduces both the compute overhead and the size of gradients for conv layers. The key contributions of this paper are:

1) We propose a novel scoring metric to model the fundamental trade-off between a block's gradient-based sensitivity and its EDP when executed on a specific architecture, called Sensitivity-EDP Ratio (SER).
2) We develop ADEPT, a framework that leverages SER to perform hardware-aware layer selection. ADEPT reduces the training complexity for fine-tuning by synergistically combining layer selection with a principled LoRA-based rank parameterization for convolutional layers.
3) We demonstrate that ADEPT achieves up to 8.1× lower EDP compared to full-parameter fine-tuning while maintaining comparable accuracy on the same PIM hardware platform. We further show that ADEPT extends naturally to continual learning by refreshing block sensitivity online at negligible overhead.

The rest of this paper is organized as follows; Section II provides relevant background. Section III outlines the problem statement, and Section IV describes the ADEPT framework. Section V presents the experimental results, and Section VI concludes the paper.

## II. Background And Related Work

In this section, we discuss PIM accelerators and provide relevant background on existing fine-tuning techniques.

### *A. PIM Accelerators*

PIM-based accelerators for Deep Neural Networks (DNNs) have been realized using various memory technologies, including CMOS-based memory devices such as Static Random-Access Memory (SRAM) and NVM technologies—including Resistive Random-Access Memory (ReRAM), Phase Change Memory (PCM), and Ferroelectric Field-Effect Transistors (FeFET) [19]. Each PIM device has its own unique advantages as well as drawbacks. NVM-based PIM architectures offer high-density matrix-vector multiplications (MVM) operations in the analog domain, making them extremely energy efficient [19], [20]. Alternatively, SRAM-based PIM offers low write latency and low device variations [21]. However, SRAM-based PIM accelerators are associated with significant area overhead and high leakage energy compared to their NVM counterparts [22]. Recently, heterogeneous integration has been proposed to address the limitations of homogeneous PIM [21], [22]. Specifically, CMOS PEs have been used for the back-propagation phase, while NVM is utilized to implement the forward phase [21], [22]. Despite these advancements, the back-propagation phase necessitates off-chip memory access on PIM devices, which becomes the primary bottleneck for training and fine-tuning.

### *B. Existing Fine-Tuning Techniques*

Several techniques exist to fine-tune DNNs efficiently. A widely adopted method for fine-tuning CNNs is transfer learning, where models are pre-trained on large-scale datasets. Afterward, only the last few layers are retrained upon deployment to adapt to specific tasks [23]. However, CNNs have most of the trainable parameters in the later layers [24]. Hence, the benefit of this approach is limited with gradient computation necessary for later layers. Gradient filtering is a recent approach that focuses on reducing the overhead of training later layers [14]. Additional techniques include gradual unfreezing of layers from top to bottom or bottom to top during fine-tuning as a form of regularization [15], [16]. The above-mentioned methods are agnostic to the data distribution. As established in recent work, such approaches may suffer from poor generalization, as certain layers might be more relevant or sensitive to specific types of data changes and prediction tasks [17], [23]. Selective fine-tuning is a state-of-the-art approach that involves determining which layers of a pre-trained model should be fine-tuned to optimize knowledge extraction [25], [26]. AutoRGN is a popular technique that dynamically modifies the learning rate of each layer based on the relative ratio of gradient norm to parameter norm [23]. AutoRGN updates all layers, similar to full-parameter training (FpT), which leads to high performance overhead. More recently, there have been PEFT methods proposed for fine-tuning transformers and foundation models. These include optimizing input layer activations, employing methods such as LoRA, and prompt tuning [27], [28], [29], [30]. However, these approaches are effective for language models that utilize text data and are less suitable for CNNs that process image data. Recent work has explored adapting LoRA to CNNs. LoRA-C applies low-rank decomposition to convolutional layers, demonstrating the feasibility of PEFT for CNNs [30]. However, LoRA-C applies a fixed-rank decomposition uniformly across all layers. Critically, none of the existing fine-tuning techniques, including LoRA-C, account for the hardware-specific training overhead when selecting which layers to fine-tune. ADEPT addresses this critical gap by co-designing the layer selection with the hardware architecture-dependent EDP cost.

## III. Problem Statement

PIM architectures excel at inference by keeping weights on-chip and reducing the costly memory access for computation. However, the fine-tuning process is significantly more memory-intensive compared to inference. PIM accelerators generally leverage pipelining across layers with mini-batches to enable high-throughput training [31], [32]. Fig. 2 presents a conceptual PIM-based Processing Element (PE) and the memory hierarchy in PIM-enabled architectures. The inputs are first loaded onto the on-chip memory of each PE. In the context of CNN fine-tuning, the images are processed in batches, with each input sent sequentially into the pipeline.

The pipelined implementation necessitates the simultaneous execution of all layers, with the weights mapped onto the crossbars. For a batch size of $B$ images, the forward pass generates a batch of $B$ output activation maps at each layer. These output activations serve as input for subsequent layers. Specifically, the back-propagation phase requires the $B$ activations generated during the forward pass. These activations are first utilized to compute the error. Next, the gradient computation necessitates access to the $B$ activation maps and the corresponding $B$ errors [13]. Once the gradients are computed for each input, the weights are updated on the PIM crossbars to process the next batch. Due to the strict area and leakage power limitations of on-chip SRAM buffers, it is infeasible to store all intermediate activations on-chip. Consequently, they are spilled to off-chip DRAM. Afterwards, the activations are fetched from DRAM based on the available on-chip memory during the backward pass. However, this off-chip data transfer is more energy-intensive and slower than on-chip computation and communication [13].

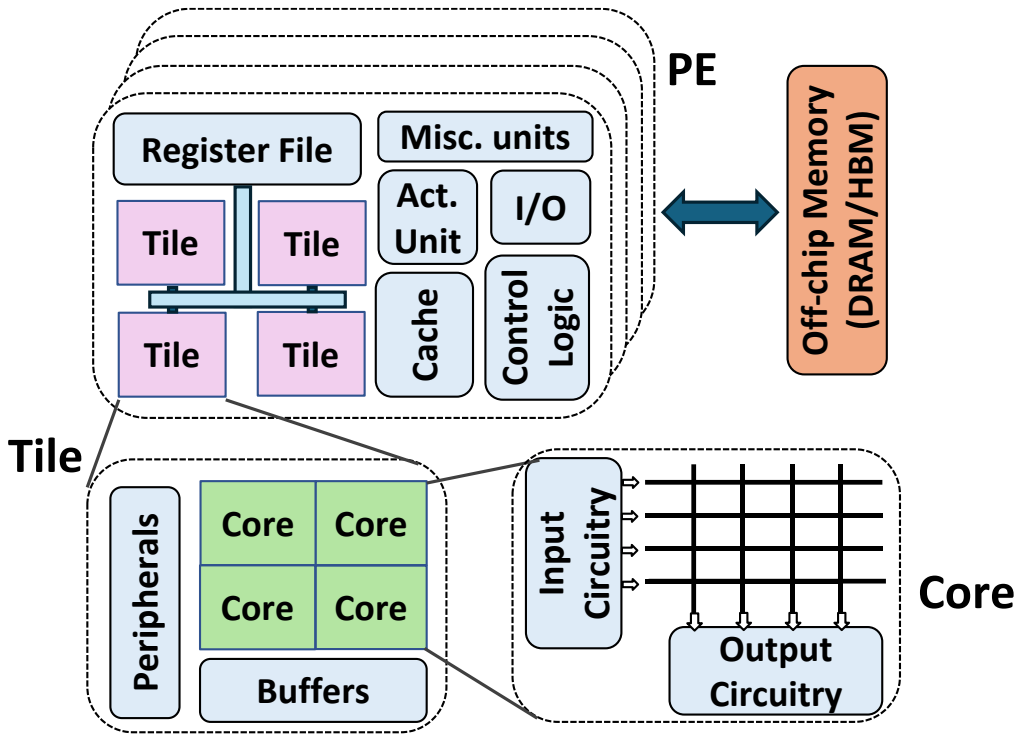

**Fig. 2:** A conceptual illustration of the PIM processing element (PE) and memory hierarchy. PIM PEs can be homogeneous or heterogeneous based on the architecture.

Existing fine-tuning techniques, as discussed in Section II, are generally oblivious to the overhead of training a layer. They treat the training overhead of all layers equally, assuming homogeneous costs. However, the neural architectural pattern of decreasing spatial dimensions while increasing channel dimensions across CNN blocks leads to non-uniform computation and communication overheads. For example, the last conv block in ResNet-18 has more model parameters than all preceding conv blocks combined [18]. This cost non-uniformity is particularly pronounced on PIM architectures since the frozen pre-trained weights already reside on-chip within the PIM arrays. Here, the layer-wise training cost is dominated by the asymmetrical off-chip DRAM access for forward activations and backward gradients. This contrasts with compute-centric platforms (e.g., GPUs), where both weights and activations are continuously fetched from DRAM. Our goal is to precisely address this gap in the current state of knowledge.

## IV. ADEPT Fine-tuning Framework

In this section, we first provide an overview of the proposed framework, ADEPT. Next, we elaborate on the hardware-aware selection criteria for fine-tuning. Afterwards, we discuss the implementation of LoRA for CNNs. Finally, we detail the end-to-end fine-tuning process and present an adaptive extension of ADEPT that sustains the SER-based selection under continual learning.

### A. ADEPT Fine-tuning

An effective fine-tuning technique must be hardware-aware, co-designing the layer-selection strategy with the overhead of training a particular layer. The ADEPT framework enables this by determining a one-time hardware-aware ranking of blocks before fine-tuning. Second, a nested block-wise training procedure is employed, which utilizes the selection order to enable architecture-aware fine-tuning.

**Hardware-aware block selection:** For a pre-trained model composed of multiple CNN blocks, our goal is to determine an optimal fine-tuning configuration that maximizes the accuracy of the network and minimizes the training cost. To this end, we define a novel metric called the *Sensitivity-EDP Ratio (SER),* as the proxy for selecting blocks during fine-tuning. *SER* explicitly measures a block's sensitivity against its training overhead. It is defined as:

$$SER_b = \frac{(S_b)^n}{(T_b)^m} \quad (1)$$

Here, $S_b$ denotes the block's sensitivity and $T_b$ quantifies its associated training overhead. The *SER* metric provides a tunable trade-off between sensitivity and cost, enabling finetuning configurations unattainable with sensitivity alone. The exponents $n$ and $m$ can be tuned to prioritize accuracy ($n > m$) or efficiency ($m > n$) depending on the target use-case application. Once $SER_b$ scores are computed for all blocks, they are sorted from the highest to lowest values. This creates a prioritized selection order $\{B_1, B_2, B_3, \dots, B_z\}$, where $B_1$ is the CNN block with the highest *SER* score.

**Nested block-wise training:** During fine-tuning, ADEPT employs a stochastic, hardware-aware nested training strategy. Rather than utilizing a static threshold to freeze layers, we construct a hierarchical collection of nested training subsets based on the pre-computed sorted *SER* scores.

$$V = \{V_1, V_2, V_3, \dots, V_z\} \quad (2)$$

Here, each subset $V_k$ contains the top $k$ highest-priority blocks, (e.g., $V_1 = \{B_1\}$, $V_2 = \{B_1, B_2\}$). For each mini-batch during fine-tuning, one subset $V_k$ is uniformly sampled from the collection $V$. The CNN blocks in the subset $V_k$ are set as trainable, and the rest of the blocks are frozen for the mini-batch. This stochastic approach effectively trains the model with varying "effective CNN blocks" simultaneously. Since $V_k$ is uniformly sampled, the probability of a block $B_i$ being updated during any given mini-batch is precisely $(z - i + 1)/z$. Crucially, this ensures that a block's update frequency is directly proportional to its hardware-aware SER priority. The highest-priority blocks (e.g., $B_1$) receive continuous gradient updates, resulting in stable loss convergence comparable to full-parameter training (FpT), while lower-priority blocks are updated proportionally less often. Consequently, blocks holding near-equal SER scores occupy adjacent ranks and receive near-equal update probabilities, differing by only $1/z$. Conversely, by stochastically freezing these lower-priority blocks, ADEPT avoids fetching their intermediate activations and computing their gradients, directly mitigating the off-chip memory bottleneck [33]. This stochastic exclusion functions as an energy-aware structured dropout. Training with this stochastic depth, where the survival probabilities of blocks are explicitly dictated by their hardware-aware SER scores, also helps improve overall model generalization [34]. Fig. 3 illustrates the fine-tuning approach employing ADEPT for a mini-batch.

*B. Sensitivity-EDP Ratio (SER)*

The *SER* for a CNN block $b$ is determined from two components: its sensitivity ($S_b$) and its training overhead ($T_b$).

**Sensitivity**: The sensitivity of a block is quantified using the relative gradient norm, a metric that reflects the block's sensitivity to the fine-tuning dataset [23]. It is calculated as:

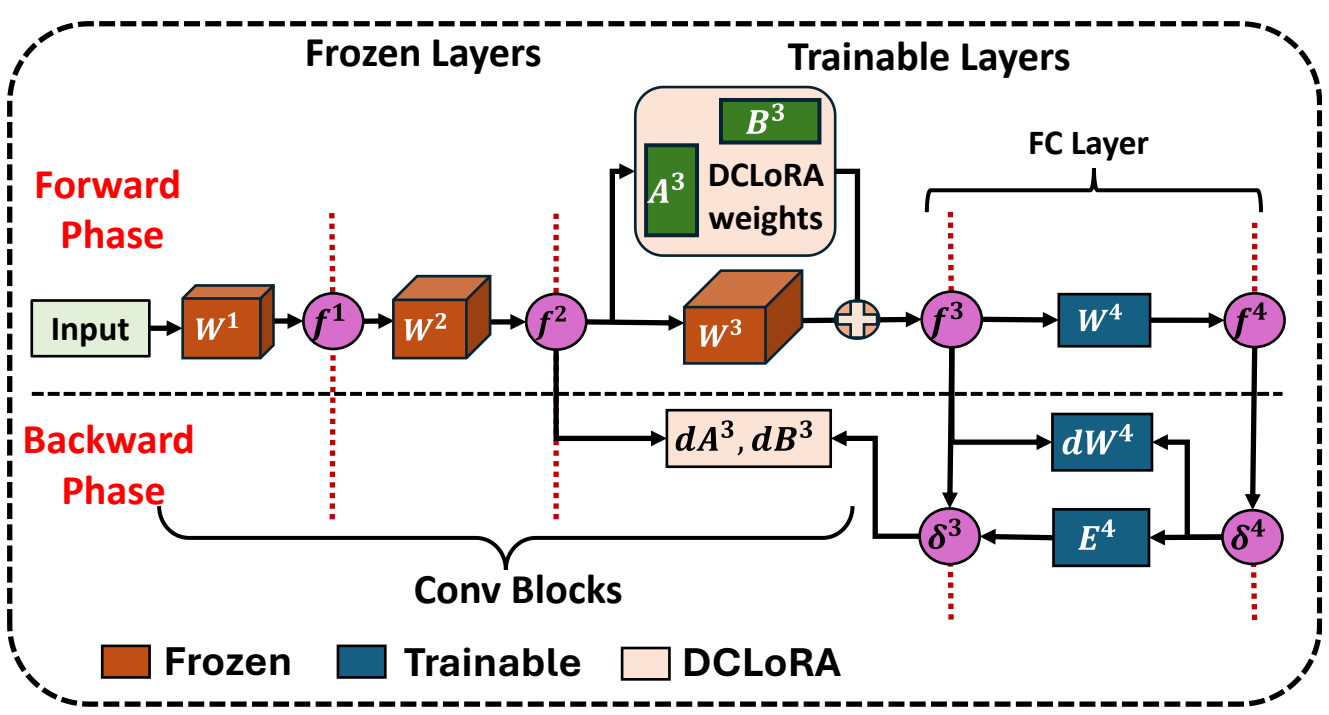

**Fig. 3:** High-level illustration of the fine-tuning pipeline employing ADEPT for a mini-batch. The back-propagation computation generates gradients only for the trainable parameters uniformly sampled from $V$. In this illustration, these include the LoRA parameters ($A^3$and $B^3$) for the last conv block and the weights ($W^4$) for the FC layer.

$$S_b = \frac{\|g_b\|_2}{\|\theta_b\|_2} \quad (3)$$

Here, $g_b$ and $\theta_b$ represent the flattened gradient and model parameters for the block $b$, respectively. The normalized values help determine the CNN blocks where we observe significant gradient change relative to the block's existing pre-trained parameters. Consequently, a high $S_b$ thus flags blocks needing substantial adaptation to capture the shifted data distribution of the fine-tuning task. Note that $S_b$ depends only on the relative order of SER scores, not on their absolute magnitudes. Therefore, the ranking is invariant to any uniform multiplicative perturbation of the training overhead, since argsort$[(S_b)^n/(c.(T_b))^m]$ = argsort$[(S_b)^n/\ (T_b)^m]$ for any common factor $c > 0$. The runtime effects that scale hardware costs symmetrically across blocks leave the selection order unchanged. Similarly, $S_b$ is a ratio (Eq. 3) rather than an absolute magnitude, mitigating its exposure to estimation noise. In scenarios when noise swaps the ranks of two near-equal blocks, the nested sampling of Section IV.A assigns them near-equal update probabilities, bounding the impact on the fine-tuned accuracy.

**Training Overhead:** The fine-tuning process involves several stages, specifically the forward pass, error computation, gradient computation, and weight update. The training overhead ($T_b$) for a block is defined as the aggregated EDP for all compute stages on the target PIM architecture. We employ EDP as the metric for training overhead since it captures both performance and energy in a single parameter.

For a given block $b$, the latency ($L$) and energy ($E$) for training can be determined using the compute and communication components. The compute latency and energy scale with the number of MVM operations, which depends upon the weights ($W^b$) and input activations ($A^{b-1}$) for block $b$. For a many-core accelerator (see Fig. 2), this can be represented as:

$$L_{compute}(b) = \alpha . MVM(W^b, A^{b-1})(L_{tile} + L_{peri}) \quad (4)$$

$$E_{compute}(b) = \beta . MVM(W^b, A^{b-1})(E_{tile} + E_{peri}) \quad (5)$$

Here, $MVM(W^b, A^{b-1})$ denotes the total scalar count of multiply-accumulate operations required for the block. $L_{tile}$ and $E_{tile}$ denote the latency and energy for computation on the PIM tile, while $L_{peri}$ and $E_{peri}$ capture the corresponding periphery circuit overheads. The constants $\alpha$ and $\beta$ are architecture-specific variables that account for utilization, tiling, and other mapping-specific overheads.

Communication modeling needs to account for both the on-chip (inter-core) and off-chip (DRAM) data exchange. For a system with $N$ PEs, these are estimated as:

$$L_{comm}(b) = L_{on\text{-}chip}(b) + \sum_{i=1}^{N} A_{Di}.(h_{mi}.L_{link} + L_{DRAM}) \quad (6)$$

$$E_{comm}(b) = E_{on\text{-}chip}(b) + \sum_{i=1}^{N} A_{Di}.(h_{mi}.E_{link} + E_{DRAM}) \quad (7)$$

where $L_{on\text{-}chip}$ and $E_{on\text{-}chip}$ represent the on-chip communication latency and energy, and $L_{DRAM}$ and $E_{DRAM}$ are the DRAM access latency and energy. $A_{Di}$ is the data volume and $h_{mi}$ represents the number of hops between DRAM and the $i^{\text{th}}$ core, respectively. Finally, $L_{link}$ and $E_{link}$ is the latency and energy consumed per bit, respectively.

We note that $T_b$ is instantiated from the actual per-block dataflow under the given weight mapping. In scenarios where a block's frozen weights cannot fully reside on-chip, the additional off-chip weight fetches are captured directly in that block's DRAM access volume ($A_{Di}$ in Eqs. 6-7), and the resulting SER scores yield a selection order consistent with the constrained mapping. Such a configuration constitutes an intermediate point between two extremes: 1) the memory-centric platform, where all frozen weights reside on-chip; and 2) the compute-centric platform, where weights are continuously fetched from off-chip memory. We characterize both these extremes in Section V (Tables IV-V, Fig. 6).

### *C. Dynamic Channel Low Rank Adaptation*

PIM accelerators face significant challenges in enabling energy-efficient back-propagation computation. The training overhead is generally dominated by the off-chip DRAM access. PEFT techniques like LoRA have achieved ubiquitous success in Transformer and generative models by uniformly decomposing large, dense linear layers. However, their straightforward application to CNNs is fundamentally limited by layer heterogeneity and spatial filter dimensions. To mitigate the off-chip DRAM access overhead during PIM-based fine-tuning, we propose a tailored, selective, and dynamic implementation of LoRA for CNNs, referred to as Channel-wise LoRA (DCLoRA).

Standard LoRA introduces additional trainable low-rank matrices [27]. Specifically, the model weights for a layer ($W \in R^{d\times d}$) are decomposed into a sum of a frozen matrix ($W_O \in R^{d\times d}$) and a product of low-rank matrices ($A \in R^{d\times r}$and $B \in R^{r\times d}$):

$$W = W_O + AB \quad (8)$$

Here, "$r$" is the rank, and "$d$" represents the dimensionality of the original weight matrix, where typically $r \ll d$. However, convolutional filters in CNNs possess small spatial dimension (e.g., $1 \times 1$ or $3 \times 3$), limiting parameter reductions achievable through standard spatial decomposition. For instance, applying the existing implementation of LoRA with a rank of 1 to a $3 \times 3$ filter results in only a 1.5× reduction in parameters.

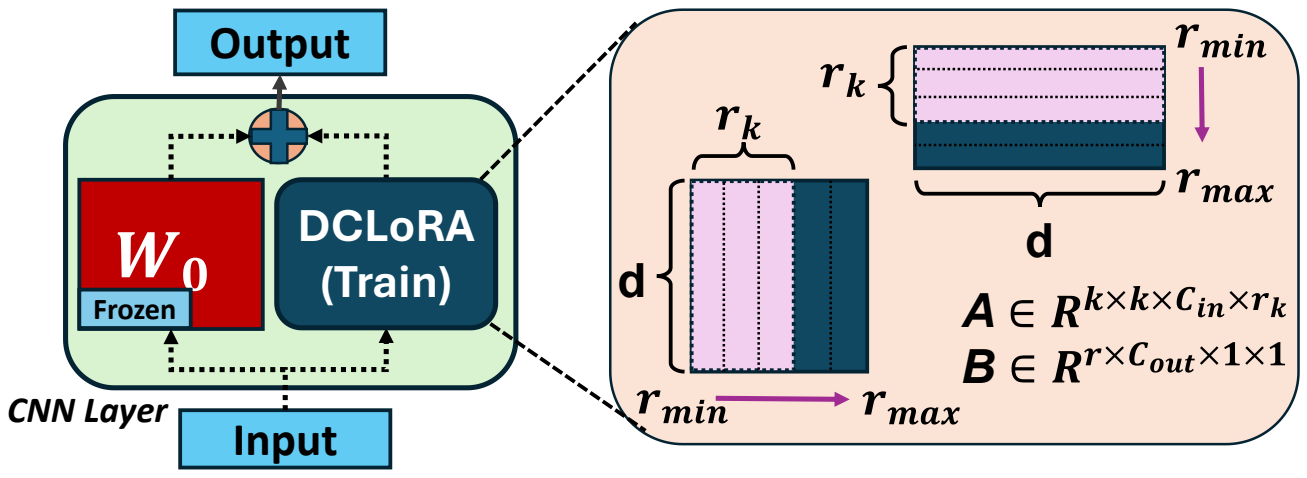

**Fig. 4**: High-level illustration of low-rank adapters added to the CNN weights. The pre-trained weights $W_O$ and LoRA matrices $(A, B)$ are mapped to PIM devices. A rank $r_k$ is sampled from the set $R = \{r_{min}, \dots, r_{max}\}$ for a block during mini-batch training.

To address this limitation, we apply *LoRA to channel dimensions instead of spatial kernel dimensions* for a CNN layer. The weights for a layer $l$ with input channels $C_{in}$, output channels $C_{out}$, and kernel size $k \times k$ can be represented as:

$$W^l \in R^{k\times k \times C_{in}\times C_{out}} \quad (9)$$

The low-rank decomposition for weights can be expressed as:

$$A^l \in R^{k\times k\times C_{in}\times r}, B^l \in R^{r\times C_{out}\times 1\times 1} \quad (10)$$

Assuming $C_{in} = C_{out} = C$ for simplification, the ratio of total trainable parameters in a conv layer without LoRA($Param_{full}$) and with LoRA ($Param_{LoRA}$) is:

$$\frac{Param_{full}}{Param_{CLoRA}} = \left(\frac{k^2 \times C \times C}{k^2 \times C \times r + r \times C \times 1^2}\right) = \left(\frac{C}{r + r/(\text{k})^2}\right) \quad (11)$$

During back-propagation, we calculate the gradient of the loss $L$ with respect to the trainable $A^l$ and $B^l$ parameters. Note that the gradients for the frozen matrix $({W_O}^l)$ are zero. The gradients $\nabla A^l$ and $\nabla B^l$ are much smaller in size than the full gradient $\nabla W^l$. This reduction in gradient size helps in decreasing the amount of off-chip data access to DRAM, which is the primary bottleneck in PIM architectures. The large, frozen weight matrix $({W_O}^l)$ can be stored on-chip within the PIM PEs. This contrasts with a typical compute-centric architecture, where both ${W_O}^l$ and the LoRA parameters are fetched from off-chip for computation.

Finally, we incorporate a dynamic rank adaptation methodology [35]. Instead of training with a fixed rank $r$, the LoRA parameters are trained across a user-specified range $R = \{r_{\text{min}}, \dots, r_{max}\}$. For each LoRA-adapted block, the full-sized adapter matrices with $r_{max}$ are initialized. Similar to the block-selection strategy detailed in Section IV.A, a nested training methodology is incorporated for rank. For each mini-batch, a rank $r_k$ is uniformly sampled from the set $R$. Fig. 4 illustrates the training methodology for nested DCLoRA training. The forward and backward passes are computed using only the truncated (nested) subsets of these matrices corresponding to $r_k$. This forces the adapter to learn an ordered representation, capturing the most critical information in the lowest-rank components, similar to structured dropout [36]. After fine-tuning, we employ the rank with highest test accuracy in $R$ for inference. Crucially, these low-rank matrices ($A^l$ and $B^l$) can be re-parameterized and merged directly into the frozen weight matrix ($W_O{}^l$), resulting in no additional latency or memory overhead during inference

### *D. End-to-End Fine-tuning Framework*

The ADEPT framework utilizes a nested training with hardware-aware layer selection and dynamic channel-wise LoRA adaptation (DCLoRA) for the selected layers. Algorithm 1 details the complete fine-tuning process. The first stage is the pre-processing step (Algorithm 1, Lines 1-7). To evaluate sensitivity for each block, we determine the flattened gradient ($g_b$) by performing a single forward and backward pass on a small subset of mini-batches. Next, we selectively initialize DCLoRA on candidate CNN blocks (Section IV.C). Unlike Transformers where LoRA is uniformly beneficial across all layers, CNNs exhibit extreme heterogeneity in their layer-wise memory bottlenecks. In early CNN layers, the training EDP is overwhelmingly dominated by the off-chip DRAM accesses required to store high-resolution forward activations, rendering any reduction in weight gradients via LoRA largely inconsequential. Conversely, the later layers possess massive channel dimensions where the weight gradients heavily dictate the off-chip memory traffic. Therefore, ADEPT applies DCLoRA *selectively*, gating its application based on a layer's specific hardware EDP reduction threshold (e.g., $2\times$ reduction in EDP compared to FpT). This selective mapping ensures that DCLoRA is only instantiated on channel-dense blocks where the reduction in gradient size meaningfully reduces the off-chip access. After DCLoRA initialization, we compute the $SER$ score for every block by evaluating its sensitivity for fine-tuning against its training cost. This scoring creates a hardware-aware priority map, which is then used to construct a collection of nested subsets of blocks ($V$).

The second stage is the nested fine-tuning process (Algorithm 1, Lines 8-15). For each mini-batch, a single subset $V_k$ is uniformly sampled from the collection $V$. Concurrently, for each LoRA-parameterized CNN block within the active subset $V_k$, a rank $r_k$ is uniformly sampled from its pre-defined range in the set $R$. This dual-level sampling dynamically determines both which CNN blocks to train and at what rank to train them if applicable. This dynamic "freeze/unfreeze" across mini-batches modifies only the gradient update computation and requires no weight movement. The complete fine-tuning loop executes on the accelerator, with the host only involved once during the pre-processing stage. Initially, the host programs the sorted priority map, the per-block DCLoRA eligibility flags, and a shared LFSR seed into per-PE configuration registers. Thereafter, no host intervention occurs for the duration of fine-tuning (see Fig. 5).

The mini-batch sampling of the subset $V_k$ and rank $r_k$ is performed natively on-chip by an LFSR-based pseudo-random generator integrated within a finite state machine (FSM)-based controller in each PE. The FSM operates with three primary states per resident block: (i) *Idle*, where the block is frozen and no gradient computation is performed; (ii) *Active-FpT*, where the block is unfrozen for full-parameter training; and (iii) *Active-DCLoRA*, where only the LoRA adapter matrices are updated at the dynamically sampled rank. Note that the per-block update frequency is determined structurally by the nested

**Algorithm 1.** ADEPT: Hardware-Aware Fine-tuning Framework

**Input**: $NN$ = Pre-Trained CNN Model with $z$ blocks, $D$ = Training Dataset, $r_{min}, r_{max}$ = rank range, $n, m$ = Score exponents, $E$ = epochs, $Tr$ = User-defined threshold for DCLoRA, $Arch$ = PIM Architecture
**Output**: $NN_{tuned}$ = Fine-Tuned CNN Model

1: *Initialize Scores*←∅, *V*←∅ // Pre-Processing step
2: **for each** $b \in blocks(NN)$ **do**
3:    *If EDP_Reduction(b) > Tr: Use DCLoRA for b*
4:    *Compute* $SER_b$*(using* Eq. 1*) and append* $[SER_b, b]$ *to Scores*
5: **for** $k = 1$ **to** $z$ **do**
6:    $V_k \leftarrow Top_k(Sorted(Scores))$
7:    $V.append(V_k)$
8: **for** *epoch* $e = 1$ **to** $E$ **do** // Nested Block-wise Training
9:    **for each** mini-batch *in D* **do**
10:       $Freeze\ (NN, Arch)$
11:       $V_k \leftarrow Sample(V)$
12:       **for each** block $b$ **in** $V_k$ **do**
13:          **if** (*is_DCLoRA(b)*): *DCLoRA* with $r_k \leftarrow Sample(R)$
14:          **else:** $Unfreeze(b, Arch)$
15:       $Train(NN, Arch)$
16:       **return** $NN_{tuned}$ // Fine-tuned model collection $V$. Each block $B_i$ is contained in every subset from $V_i$ to $V_z$, guaranteeing that a uniform draw over the $z$ subset indices updates $B_i$ with a fixed frequency of $(z-i+1)/z$. State transitions occur strictly at mini-batch boundaries, driven by the decoded training subset of active blocks ($V_k$) and the adapter rank ($r_k$). Crucially, this boundary already constitutes the pipeline-drain point at which the pipeline is cleared to perform weight updates on pipelined PIM accelerators [31], [32]. ADEPT's state transitions are aligned to this pre-existing barrier. Overall, the framework introduces no additional synchronization point and no deviation from the baseline memory-access pattern beyond the intended elimination of fetches for frozen blocks.

Within each mini-batch, the FSM asserts control signals to the PE's compute units and configures the DRAM fetch logic to request only the data required by the active training mode, thereby explicitly circumventing the severe off-chip memory accesses for frozen blocks. Synthesis results show that this control logic, including the LFSR and configuration registers, introduces less than 0.5% tile area and power overheads at 32 nm technology node. This overhead is negligible compared to the EDP savings achieved using ADEPT, as we show later.

### *E. Extension to Continual Learning*

Continual learning deployments frequently encounter non-stationary data distributions, which necessitates periodic adaptation [8]. However, repeated execution of the calibration pass, as detailed in the previous subsection, for every distribution shift would impose a recurring profiling overhead that significantly degrades the overall EDP. To address this challenge, ADEPT leverages the structural decomposition of the SER metric to enable an online tracking mechanism, as illustrated by the continual learning feedback loop in Fig. 5. As defined in Eq. 1, SER decomposes into a hardware-specific training overhead ($T_b$) and a data-dependent sensitivity ($S_b$). $T_b$ is derived from the per-block dataflow and the EDP characterization of the target platform (Eqs. 4-7). Therefore, it is fixed once the CNN model, and the associated hardware platform are specified and remains strictly invariant across fine-tuning tasks. Consequently, only $S_b$ drifts under distribution shifts, necessitating it to be refreshed to sustain an up-to-date block prioritization.

To mitigate the off-chip memory wall associated with continuous re-profiling, ADEPT refreshes $S_b$ online using an Exponential Moving Average (EMA) computed entirely from the gradients inherently produced during the backpropagation phase. For each mini-batch, the backward pass generates the gradients of the trainable parameters for every active block $b$ within the stochastically sampled subset $V_k$. These gradients yield a new sensitivity sample $x_b$ on the current mini-batch, which updates the running estimate as:

$$S_b \leftarrow S_b + (1 - \lambda)\,(x_b - S_b) \qquad (12)$$

Here, $\lambda \in (0,1)$ is the smoothing factor, with $S_b$ initialized during the one-time calibration step in the pre-processing phase. Note that Eq. 12 is algebraically identical to the conventional EMA recursion, $\lambda \cdot S_b + (1-\lambda) \cdot x_b$. Crucially, this error-feedback form exposes the entire update as exactly one subtraction followed by one fused multiply-accumulate (MAC) per block. To further optimize execution and bypass the need for a multiplier, the smoothing factor is set to be $\lambda = 1-2^{-k}$, where k is a positive integer. This makes the weight on the new sample, $(1-\lambda) = 2^{-k}$, a power of two, so the update $S_b \leftarrow S_b + 2^{-k}(x_b - S_b)$ reduces to a subtraction, an arithmetic right-shift by k bits operation, and an addition, requiring no multiplier.

All operands necessary to compute EMA (Eq. 12) are produced natively on-chip, explicitly circumventing any additional off-chip DRAM accesses. During the weight-update stage, each parameter-gradient pair $(\theta_i, g_i)$ already streams to the PE from off-chip memory (Eq. 3), imposing no additional memory fetches. At the mini-batch boundary, where the pipeline is already drained to synchronize weight updates, the per-PE partial sums are communicated over the on-chip interconnect. However, it constitutes negligible traffic relative to the weight-update dataflow. The resulting per-block scalar ratio and the subsequent EMA update execute directly on the Special Function Units (SFU)/Activation unit within the PE. These functional units already provide the element-wise arithmetic required for activation functions, error computation, and occupy under 0.1% of chip area and power in existing PIM accelerators [22], [37], [38]. Therefore, ADEPT essentially repurposes a negligible fraction of the computation on existing hardware substrate, necessitating no new functional units or complex control logic to support continual learning.

The updated $S_b$ values are subsequently combined with the static $T_b$ constants to determine SER (Eq. 1) and update the priority map at a configurable interval aligned with mini-batch boundaries (epoch boundaries in our evaluation). Since this state transition strictly aligns with the pre-existing mini-batch boundaries used for weight-update pipeline drains, it introduces no new inter-PE synchronization barriers. The updated block priorities are written to the per-PE configuration registers described in Section IV.D, necessitating no host CPU intervention. The DCLoRA eligibility decisions remain unchanged, as the gating threshold depends solely on the hardware-determined EDP reduction and is independent of $S_b$.

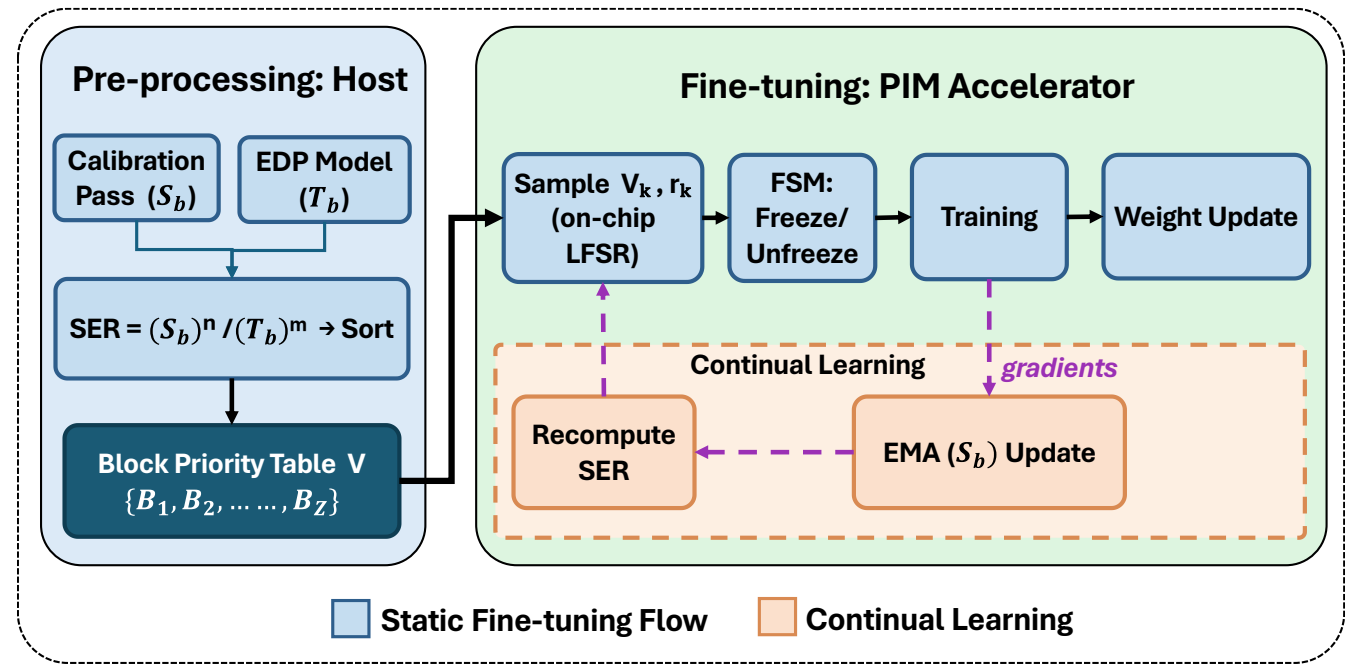

**Fig. 5**: Execution flow of the ADEPT framework. The host performs a one-time calibration pass to initialize the training. During fine-tuning, the PIM accelerator orchestrates stochastic sampling and state transitions. For continual learning deployments, ADEPT introduces an on-chip tracking loop (dashed) that updates the sensitivity ($S_b$) via EMA, enabling dynamic SER re-evaluation without host intervention.

The proposed scheme maintains a single running scalar $S_b$ across mini-batches, yielding a constant O(1) memory footprint over the deployment lifetime. As discussed in Section IV.A, every block is sampled with non-zero probability $(z-i+1)/z$ and updated multiple times within an epoch. Since $S_b$ is a single scalar rather than an accumulated sample history, a single refresh during an epoch is sufficient to register a block whose sensitivity has changed under a distribution shift. Consequently, once the priority map is recomputed at the epoch boundary, the blocks with higher SER value are sampled more frequently in the subsequent epoch. We empirically validate this adaptation mechanism on a sequential, multi-stage continual-learning corruption stream in Section V.F and compare it with full-parameter training.

## V. Experimental Results

In this section, we first describe the setup used for the performance evaluation. Next, we analyze the efficiency of channel-wise LoRA and the proposed layer selection process. We then present a hyperparameter sensitivity analysis, followed by an assessment of ADEPT with respect to existing fine-tuning techniques in terms of predictive accuracy and performance.

TABLE I
CNN Models and Hyperparameters

| CNN Models | Architecture | | Hyperparameters |
|---|---|---|---|
| | Conv Blocks | FC Layers | lr = [0.05, 0.01, 0.001, 0.005] |
| VGG-16 | 5 | 3 | B = 64, E=100 |
| ResNet-18 | 4 | 1 | B = 128, E=100 |
| MbNet-V2 | 17 | 1 | B = 128, E=100 |
| SnetV2-x1 | 5 | 1 | B = 128, E=100 |

TABLE II
CNN Pre-Training Model Accuracy

| | VGG-16 | ResNet-18 | MbNet-V2 | SnetV2-x1 |
|---|---|---|---|---|
| CIFAR10 | 95.9% | 94.90% | 93.6% | 94.5% |
| CIFAR10-c | 74.5% | 75.1% | 74.8% | 75% |
| Entity-30 | 74% | 75.2% | 76% | 71% |
| ImageNet | 74.8% | 73.40% | 74% | 73.9% |

TABLE III
ARCHITECTURAL SPECIFICATIONS OF EVALUATED PIM PLATFORMS

| Parameter | PUMA | CIMAT | HuNT |
|---|---|---|---|
| Memory | ReRAM PIM | SRAM PIM | (FeFET+ReRAM+ SRAM) PIM |
| Integration | 2D planar | 2D planar | 3D stacked, 4 tiers |
| Crossbar/array size | 128×128 | 128×128 | ReRAM- 128×128, SRAM/FeFET - 256 ×256 |
| On-chip organization | 138 tiles/PE, 8 cores/tile | 9 tiles/PE, 16 cores/tile | 4 tiers × 16 PEs/tier |
| Activation unit (SFU/VFU) | SFU:0.03%, VFU: ~4.9% | SFU:<0.1%, Global Buffer: 0.47% | SFU:<0.1%, Global Buffer: 0.47% |

### *A. Experimental Setup*

**CNN models and datasets:** To evaluate the efficacy of the proposed framework, we consider four diverse models: VGG-16, ResNet-18, ShuffleNetV2-x1 (SnetV2-x1), and MobileNet-V2 (MbNet-V2) [12], [18], [39]. The evaluation is conducted across four datasets: CIFAR10, CIFAR10-c (corrupted), Entity-30, and ImageNet [40]. These datasets and models represent the typical size and diversity of use cases for deployment at the edge. Table I presents the details of the datasets, and hyperparameters such as batch size (B) and training epochs (E). We consider several learning rates (lr) and report the best model accuracy for each dataset. The datasets CIFAR10 and ImageNet are partitioned into two subsets that are non-independent and identically distributed (non-IID) following the federated averaging method [7]. The non-IID partitioning ensures that the data distributions across the subsets are distinct. For instance, in the ImageNet dataset, the pretraining and fine-tuning partitions overlap on 99 out of 1000 classes (less than 10%). This helps simulate scenarios where pre-trained models encounter new data distributions. In our setup, one partition is utilized for pre-training and the other for fine-tuning. The other datasets include different types of distribution shifts. For CIFAR10-c, the model is pre-trained on the standard CIFAR10 dataset, while the fine-tuning data consists of corrupted images [23]. For Entity-30, the pre-trained model and the fine-tuning dataset share the same output classes, but the fine-tuning data contains different subpopulations of those classes [23]. Table II shows the accuracy of the models considered in our evaluation on the pre-training partition.

**Hardware and ADEPT Setup:** We employ a modified NeuroSim v2.1 to obtain the area, latency, energy of PIM tiles, all on-chip buffers, and peripheral circuits [13]. NeuroSim has been extensively validated against fabricated PIM test chips and is widely adopted as the benchmarking framework for PIM accelerators [13]. The activation and miscellaneous vector units are characterized using publicly available per-unit area/energy breakdowns of the respective platforms. For instance, PUMA reports the special functional unit (SFU) occupies under 0.1% of chip area and power [38]. Conversely, the Vector Functional Unit (VFU) in PUMA is used for GEMV and element-wise operations. It occupies 4.9% of the chip area in PUMA [38]. The SFU in CIMAT and HuNT utilizes NeuroSim and report it as a negligible fraction of chip area and power. Specifically, The activation unit and max-pooling units occupy 0.04% and 0.03% of the chip area, respectively and the global buffers account for 0.47% [13], [21], [37]. Table III summarizes the architectural parameters of the evaluated platforms.

The control logic in ADEPT fall outside NeuroSim's scope and is characterized as follows. The FSM-based controller is synthesized at the 32 nm node using the *Synopsys Design Compiler*. Additionally, the power is calculated via *Synopsys Spyglass* consuming 0.5% of tile area and power. For continual learning scenario, the online sensitivity update reuses the SFU already present in the PE and introduces no new functional unit. Consequently, no additional hardware characterization is required beyond the FSM control logic synthesized above. An industry-standard High Bandwidth Memory (HBM2E) is used as the DRAM for off-chip data access [41], consistent with the memory interface considered by the existing PIM accelerators [21], [38]. The identical DRAM parameters are applied to ADEPT and every baseline fine-tuning method. Therefore, the reported gains arise solely from ADEPT's reduction in the volume of off-chip activation and gradient traffic, not from the DRAM technology choice. We further quantify this independence with an LPDDR5X-based evaluation later in Section V.E.

We assume that the pre-trained model parameters are mapped to PEs before fine-tuning in our performance evaluation following the baseline mapping. Additionally, we take into account the overhead associated with loading activations from the DRAM for reporting performance and energy metrics. A 16-bit fixed-point precision was utilized for the computation and storage of the model parameters and activations. ADEPT employs DCLoRA selectively if a block's EDP reduction compared to FpT exceeds a user-specified threshold ($Tr$). We set $Tr = 2\times$ as a conservative threshold in our evaluation to ensure we employ DCLoRA where it leads to significant EDP savings for introducing new parameters. For layers employing DCLoRA, we consider ranks with a range of $r_{min} = 8$ and $r_{max} = 64$ in multiples of 2, where $R = \{8,16,32,64\}$ [37]. Additionally, we utilize $n = m = 1$ as hyperparameters for our $SER$ computation, balancing sensitivity, and training cost equally. We perform a hyperparameter sensitivity analysis later in Section D. To ensure fairness, ADEPT and all baseline methods are trained for the exact same total number of epochs.

### *B. ADEPT LoRA Evaluation*

In this section, we analyze the variation in per-block EDP with and without the use of DCLoRA while fine-tuning. Fig. 6 illustrates the normalized EDP for fine-tuning each conv block of ResNet-18 using FpT and DCLoRA. We model the EDP for two distinct architectures: a memory-centric PIM platform (PUMA [38]) and a compute-centric platform (TPU) [42]. PUMA is a ReRAM-based PIM accelerator featuring 128×128 crossbar configurations [38]. It comprises 138 tiles per PE, each containing 8 cores per tile.

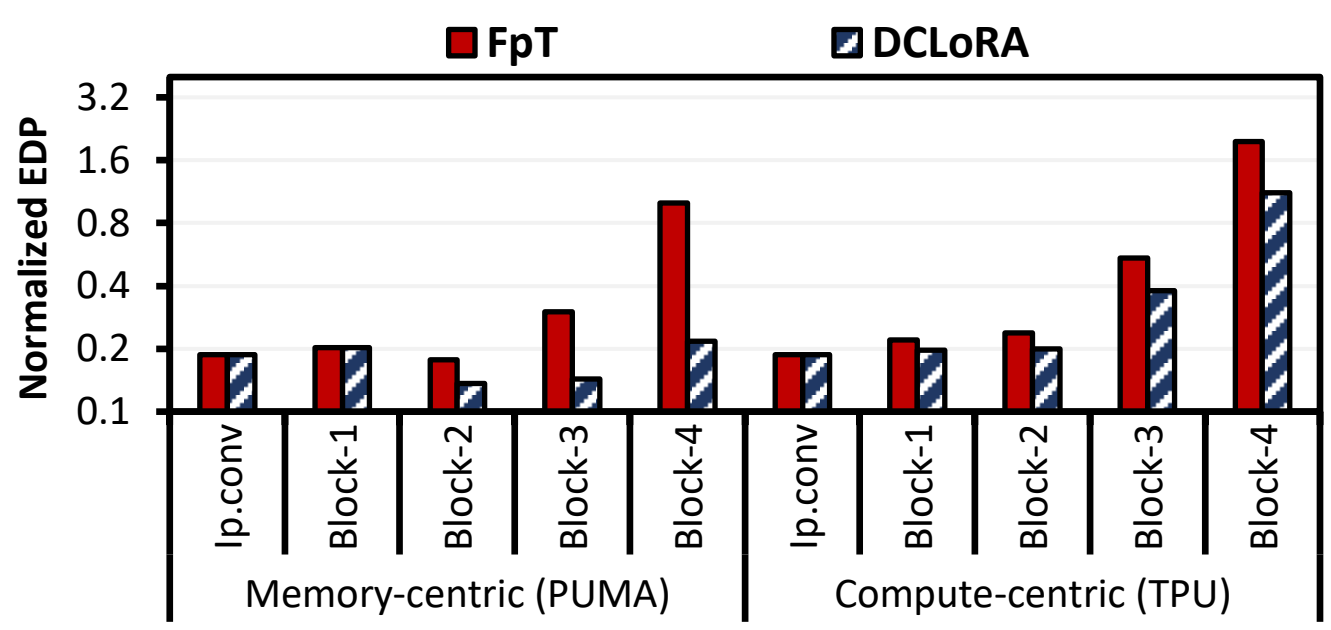

**Fig. 6:** Per-block EDP comparison of FpT and DCLoRA across conv blocks of ResNet-18, normalized independently for each architecture to the EDP of Ip.conv on that platform.

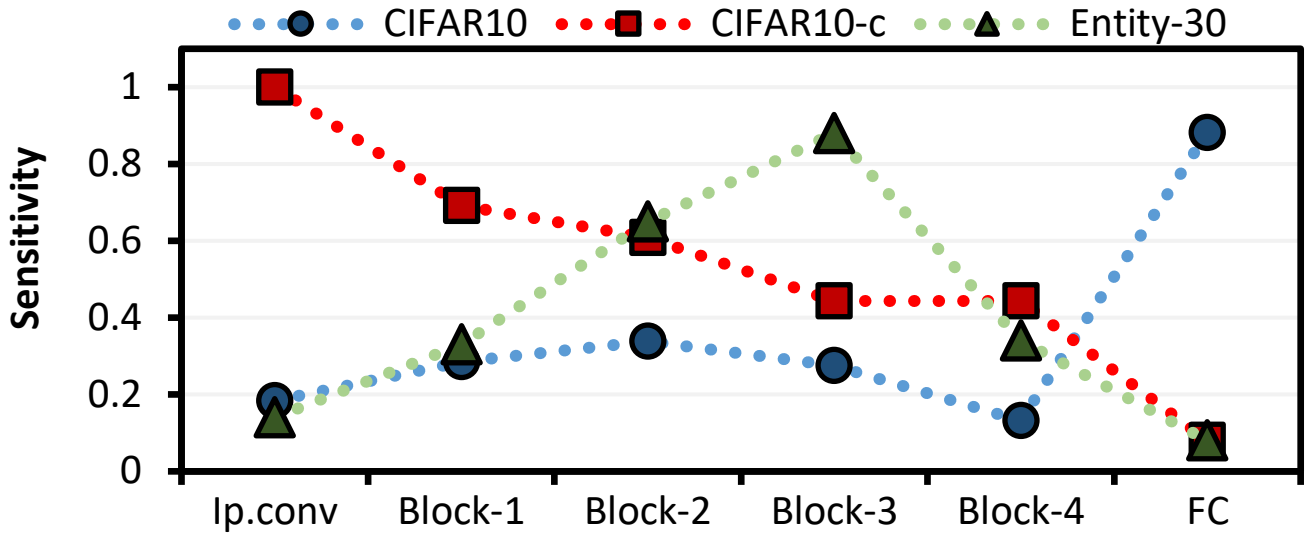

**Fig. 7:** Comparative analysis of block sensitivity ($S_b$) considering CIFAR10, CIFAR10-c and Entity-30 fine-tuning datasets in ResNet-18 (see Eq. 3).

From Fig. 6, we see little to no benefit of applying DCLoRA to initial layers (Ip.conv, Block-1, Block-2). This is because the training cost in these blocks is dominated by the large forward activations that must be stored for the backward pass. As detailed in Section IV.C, LoRA helps only in reducing the size of gradients and weight update parameters. However, for PUMA, applying DCLoRA to the latter, channel-dense blocks yields a significant EDP reduction. For instance, in Block-4, DCLoRA reduces the EDP by $\sim 4\times$ compared to FpT. This gain is because the memory-centric architecture stores the large frozen weights on-chip. The off-chip memory access is required to store the gradients for the trainable DCLoRA parameters and forward activations, which are both relatively smaller in size for Block-4. In contrast, the compute-centric TPU architecture shows lower benefit in EDP for later layers with DCLoRA. This is because it stores model parameters in DRAM and needs memory access for both DCLoRA and frozen weights.

### C. ADEPT Sensitivity Analysis

We now illustrate the model sensitivity per block for a CNN and the corresponding change in our *SER* score values based on the fine-tuning task. Fig. 7 shows the sensitivity of each block in ResNet-18 on three fine-tuning datasets: CIFAR10, CIFAR10-c, and Entity-30 as an example. We see that the sensitivity varies significantly based on the distribution shift in the fine-tuning dataset. For CIFAR10, the later layers hold higher sensitivity, aligning with common heuristics employed in the transfer learning literature. Conversely, for the corrupted CIFAR10-c dataset, the initial layers become significantly more important, as the model must adapt to low-level feature corruptions. This result empirically confirms that static fine-tuning strategies, such as only training the last few layers, are not generalizable with suboptimal performance.

Tables IV and V show the resulting block selection order based on the *SER* scores for the memory-centric PIM (PUMA) and compute-centric (TPU) platform, respectively. For PUMA (Table IV), the FC layer is ranked as the highest priority for all datasets. This is a consequence of the *SER* metric: the FC layer in ResNet-18 has limited output activations (no. of classes), resulting in minimal training cost ($T_b$). After the FC layer, the block order varies significantly based on the dataset-specific sensitivity values, as seen in the difference between the CIFAR10 and CIFAR10-c. In contrast, TPU (Table V) shows a different prioritization. Block-4 is ranked last for all datasets, even with the use of DCLoRA. This is due to the extremely high cost ($T_b$) of fetching and computing its large frozen parameters from off-chip memory, unlike PUMA. Further, we see that for CIFAR10-c, Block-3 is prioritized ahead of Block-4, despite having similar sensitivity values. However, this is not the case for CIFAR10-c on PUMA. ADEPT through the proposed *SER* metric creates different layer-selection priorities depending on the target architecture (memory-centric vs compute-centric), which prior methods fundamentally cannot capture.

TABLE IV
SER-GUIDED BLOCK SELECTION FOR MEMORY-CENTRIC (RESNET-18)

| | | | | | | |
|---|---|---|---|---|---|---|
| **CIFAR10** | FC | Block-2 | Block-1 | Ip.Conv | Block-3 | Block-4 |
| **CIFAR10-c** | FC | Ip.Conv | Block-2 | Block-1 | Block-4 | Block-3 |
| **Entity-30** | FC | Block-2 | Block-3 | Block-4 | Ip.Conv | Block-1 |

TABLE V
SER-GUIDED BLOCK SELECTION FOR COMPUTE-CENTRIC (RESNET-18)

| | | | | | | |
|---|---|---|---|---|---|---|
| **CIFAR10** | FC | Block-1 | Block-2 | Ip.Conv | Block-3 | Block-4 |
| **CIFAR10-c** | Ip.Conv | Block-1 | Block-2 | Block-3 | FC | Block-4 |
| **Entity-30** | Block-2 | Block-3 | Block-1 | Ip.Conv | FC | Block-4 |

### D. ADEPT Hyperparameter Analysis

We now analyze the sensitivity of ADEPT's hyperparameters: the exponents $n$ (sensitivity) and $m$ (training overhead). As detailed in Section IV, these exponents dictate the weightage of the SER components and can be adjusted to prioritize either predictive accuracy ($n > m$) or hardware efficiency ($m > n$) based on the target use-case.

Table VI presents the block-selection order of ResNet-18 on CIFAR10 for PUMA, sweeping $n$ and $m$ across {0, 1, 2} as an example. The results show that the block-selection order for our nested training changes based on the chosen exponents. Fig. 8 illustrates the corresponding impact on overall model accuracy and training EDP. The boundary cases where $n = 0$ or $m = 0$ represent scenarios where the layer-selection objective is driven entirely by EDP or sensitivity, respectively. We observe that setting $n = 0$ (oblivious to sensitivity) represents a policy that prioritizes blocks solely by their training cost and strictly selects low-cost blocks during fine-tuning. This hardware-aware selection heuristic severely limits the model's ability to adapt, leading to a significant reduction in accuracy (see Fig. 8). Therefore, EDP awareness alone is therefore insufficient, underscoring why combining sensitivity and cost into a single metric is essential. In contrast, setting $m = 0$ (oblivious to hardware cost) relies solely on sensitivity, resulting in a 70% increase in EDP compared to the balanced baseline ($n = m = 1$). Additionally, the intermediate tuning of these parameters allows for fine-grained trade-offs. Setting $m > n$ (e.g., $m = 2$, $n = 1$) penalizes high training costs, which slightly degrades accuracy by under-training sensitive but computationally expensive layers. Conversely, setting $n > m$ (e.g., $n = 2, m = 1$) prioritizes highly sensitive blocks, mimicking a sensitivity-only selection and sacrificing the EDP benefits of hardware awareness. Overall, the configuration with $n = m = 1$ can be used to equally prioritize predictive performance and training overhead, confirming its robustness. We utilize this balanced configuration ($n = m = 1$) for all subsequent end-to-end performance evaluations as the default setting in ADEPT. We further note that the SER exponents $(n, m)$ are static, one-time design choices configured before fine-tuning and the balanced setting ($n = m = 1$) is robust across the configurations studied.

TABLE VI
BLOCK SELECTION WITH VARYING HYPERPARAMETER ($n$ AND $m$)

| | | | | | | |
|---|---|---|---|---|---|---|
| [n=1,m=0] | FC | Block-2 | Block-3 | Block-1 | Ip.Conv | Block-4 |
| [n=2,m=1] | FC | Block-2 | Block-1 | Ip.Conv | Block-3 | Block-4 |
| [n=1,m=1] | FC | Block-2 | Block-1 | Block-4 | Ip.Conv | Block-3 |
| [n=1,m=2] | FC | Block-2 | Block-1 | Block-3 | Ip.conv | Block-4 |
| [n=0,m=1] | FC | Block-4 | Block-2 | Ip.Conv | Block-1 | Block-3 |

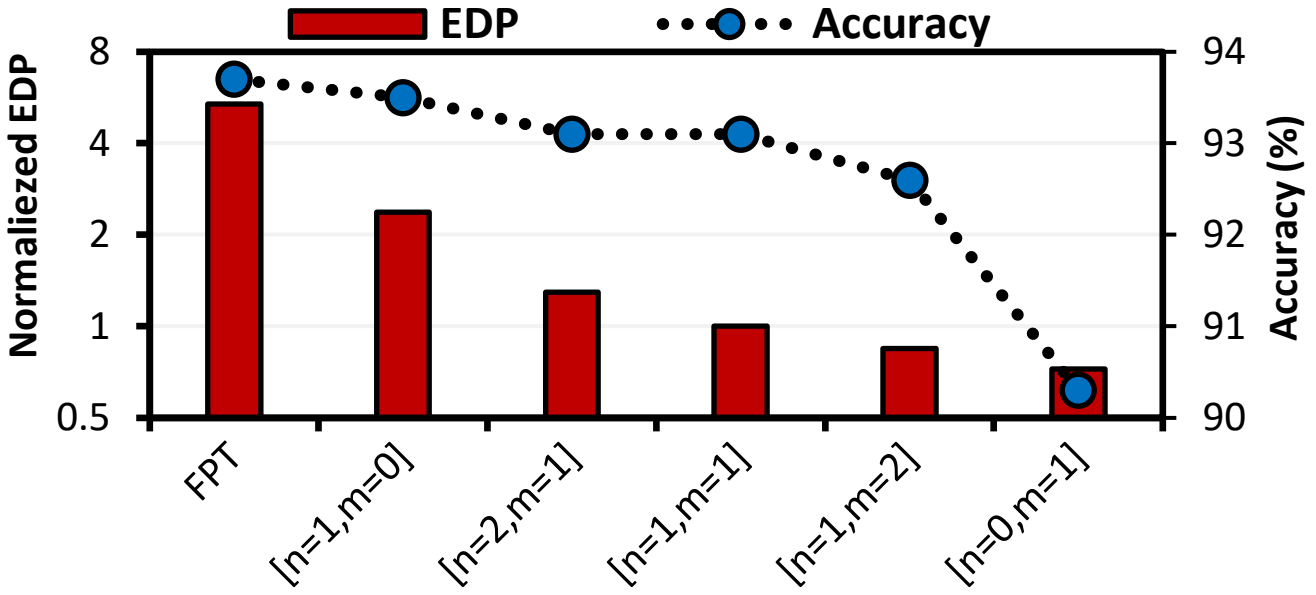

**Fig. 8:** Trade-off between predictive accuracy and normalized training EDP across different SER hyperparameter configurations ($n$ and $m$) for ResNet-18 on CIFAR10.

TABLE VII
CNN MODEL ACCURACY

| | ResNet-18 | | | | MbNet-V2 | | | | SnetV2-x1 | | | |
|---|---|---|---|---|---|---|---|---|---|---|---|---|
| **Methodology** | **CIFAR10** | **CIFAR10-c** | **ImageNet** | **Entity-30** | **CIFAR10** | **CIFAR10-c** | **ImageNet** | **Entity-30** | **CIFAR10** | **CIFAR10-c** | **ImageNet** | **Entity-30** |
| No Fine-tuning | 17.0% | 47.6% | 4.1% | 52.6% | 17.8% | 40.8% | 4.7% | 49.2% | 15.8% | 35.8% | 4.1% | 49% |
| GF | 92.8% | 69.8% | 68.6% | 67% | 92.9% | 61% | 63.4% | 64.2% | 92.3% | 61.9% | 63.3% | 63.2% |
| AutoRGN | 93.6% | 75.3% | 72.2% | 74.9% | 93% | 76.4% | 67.5% | 73.2% | 92.4% | 77.6% | 67.1% | 71.7% |
| L2F | 93.4% | 71.8% | 72.2% | 72.1% | 93.2% | 72.8% | 67.7% | 67.8% | 92.7% | 71.9% | 67.2% | 68.1% |
| ADEPT(S) | 93.5% | 75.8% | 72.1% | 74.5% | 93.3% | 76.8% | 67.8% | 72.7% | 92.8% | 77.7% | 67.2% | 71.8% |
| ADEPT | 93.05% | 75.5% | 71.9% | 74.2% | 93.4% | 76.6% | 67.6% | 72.6% | 92.7% | 77.4% | 67.1% | 71.4% |

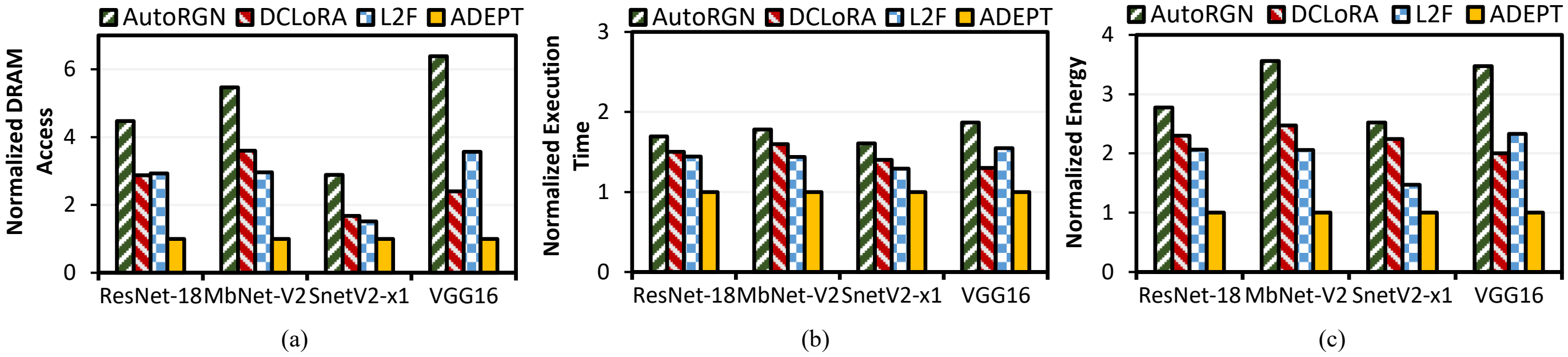

**Fig. 9:** Normalized activations size, latency, and energy of baseline fine-tuning techniques relative to ADEPT on different datasets on the PUMA architecture.

*E. ADEPT Performance Analysis*

In this subsection, we evaluate the predictive accuracy and fine-tuning performance of ADEPT and compare it against several state-of-the-art fine-tuning techniques. These include AutoRGN [23], Last-to-First (L2F) [15], Gradient-Filtering (GF) [14]. AutoRGN employs a sensitivity function to tune the learning rate of each layer while training all layers simultaneously. L2F gradually increases the number of trainable conv blocks during the fine-tuning process from last to first [23]. GF freezes earlier layers and trains only the last few layers, reducing the number of unique elements in gradient computation by spatially patching the gradient map [14]. Lastly, we include ADEPT without any hardware awareness (e.g., $n = 1, m = 0$), referred to as ADEPT(S), which uses only the sensitivity of a block for selection to show the utility of hardware awareness and DCLoRA.

**Predictive Accuracy:** Table VII presents the post fine-tuning accuracy of the ResNet-18, MbNet-V2, and SnetV2-x1 as examples employing ADEPT and the baselines. Additionally, we also report the accuracy of the pre-trained model without fine-tuning, referred to as "No Fine-tuning." From Table VII, it is evident that there is a need for fine-tuning, as the pre-trained model exhibits poor model accuracy (e.g., ~4% with the ImageNet dataset) on the fine-tuning dataset. We observe that static or heuristic-based techniques like L2F and GF struggle significantly with datasets like CIFAR10-c and Entity-30, where low-level feature adaptation is crucial. AutoRGN, through its dynamic learning rate, achieves suitable results for all datasets. However, it updates all layers simultaneously similar to FpT, making it computationally prohibitive. In contrast, ADEPT through its dataset-aware nested training methodology, achieves similar accuracy to that of AutoRGN across all models considered in this work. Notably, ADEPT's stochastic freezing does not degrade convergence as it reaches this target accuracy within the same number of epochs.

**Intermediate Activations Overhead:** Next, we compare the activations generated by ADEPT and the baselines. As established, the off-chip access is the primary bottleneck for fine-tuning on PIM architectures. The mini-batch training performs the following computations for processing each batch: forward pass, error calculation, gradient computation, and weight update. To enable backpropagation, activations from all forward layers must be stored alongside errors and gradients for the complete batch. Fig. 9(a) presents the normalized intermediate activations size during the fine-tuning process for ADEPT and the baselines L2F, AutoRGN. We also consider a naive "DCLoRA" baseline, where we apply DCLoRA on all layers similar to the conventional use of LoRA in other DNNs and transformer architecture. We exclude Gradient Filtering (GF) from this evaluation, as it cannot be generalized for all fine-tuning tasks and suffers from poor accuracy for Entity-30 and CIFAR10-c.

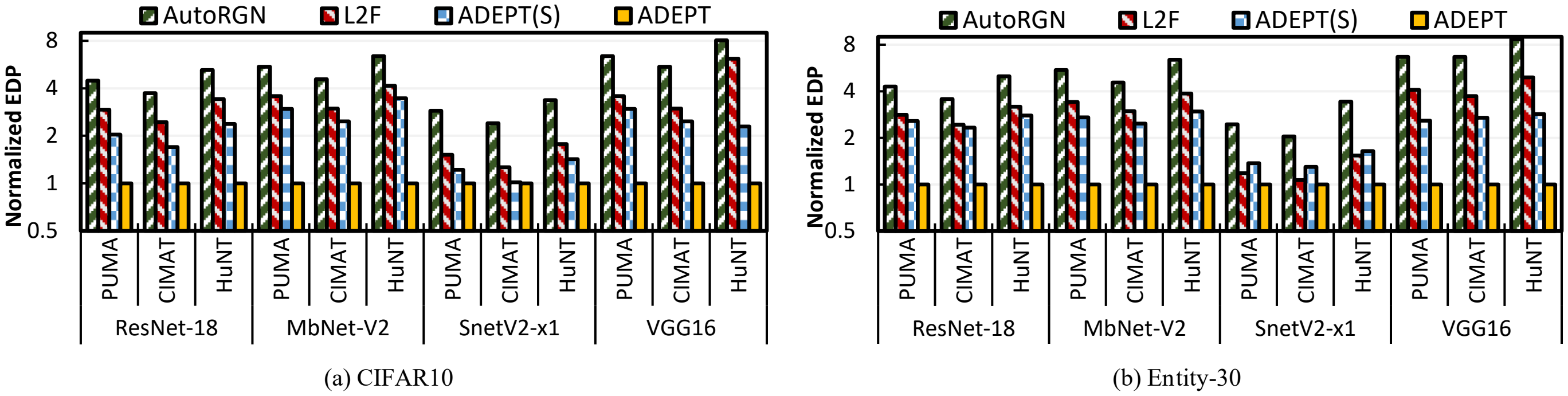

**Fig. 10:** Normalized EDP of ADEPT and baseline fine-tuning techniques on different PIM platforms with the (a) CIFAR10 and (b) Entity-30 datasets.

From Fig. 9(a), we observe that L2F reduces intermediate activations by freezing conv blocks, thereby eliminating the need to fetch forward activations for these blocks during backpropagation phase. Conversely, uniformly applying DCLoRA to all layers yields minimal reduction. This limited reduction is attributed to the dependency on channel count ($C$) when applying DCLoRA, as shown in Fig. 6. The number of channels for initial conv blocks is small. For instance, $C = 3$ for the first convolution layer, whereas $C = 1280$ for the last convolutional layer in MbNet-V2. In contrast, ADEPT significantly reduces the size of intermediate activations by synergistically combining dynamic block selection (freezing unselected layers) and selective application of DCLoRA (compressing channel-dense layers). Specifically, ADEPT reduces the intermediate-activation footprint by up to 6.38× relative to AutoRGN and up to 3.6× relative to L2F across the evaluated models (Fig. 9(a)).

**Execution Time and Energy:** Next, Fig. 9(b) and Fig. 9(c) present the normalized execution time and energy, respectively for ADEPT relative to the baseline methods. Building on the activation footprint results, applying DCLoRA uniformly to all layers provides minimal performance gains, as gradients must still be computed and activations fetched for the entire network during backpropagation. Conversely, ADEPT demonstrates substantial speedup and energy efficiency gains compared to AutoRGN.

**Cross-Platform EDP Analysis:** Finally, we evaluate the fine-tuning EDP of ADEPT and the baselines using a few representative PIM-based accelerators designed for CNNs. We consider PUMA, CIMAT, and HuNT for this analysis (see Table III) [21], [37], [38]. CIMAT proposes a SRAM-based accelerator that utilizes transposable SRAM arrays of size 128×128 [37]. In contrast, HuNT adopts a heterogeneous architecture that uses FeFET, ReRAM, and SRAM in a 3D architecture [21]. While these architectures differ in their memory technologies and configurations, they share a similar on-chip memory hierarchy, as shown in Fig. 2. We adopt ADEPT and other fine-tuning techniques on each accelerator. To ensure consistency, we maintain the same DRAM data access and energy parameters across our evaluation.

Fig. 10(a) and Fig. 10(b) illustrate the normalized EDP for AutoRGN, L2F, ADEPT(S), and ADEPT for models detailed in Table I with CIFAR10 and Entity-30 datasets as representative examples. The results for each architecture are normalized to ADEPT on the same hardware platform. We observe that ADEPT outperforms AutoRGN, ADEPT(S), and L2F in terms of EDP. ADEPT demonstrates significant gains on all the PIM platforms considered, achieving up to 8.1× reduction in EDP compared to AutoRGN. Additionally, we see a clear benefit of hardware awareness since ADEPT outperforms ADEPT(S) in terms of EDP, while ensuring effective model fine-tuning across all PIM platforms. This highlights that the SER-based selection generalizes as a plug-in enhancement to any block-level fine-tuning technique. Finally, the EDP gains vary for different datasets. This is due to the dataset and model-specific layer selection process employed in ADEPT. The primary bottleneck in PIM architectures arises from the intermediate activations generated during fine-tuning. ADEPT is precisely designed to address the frequent off-chip memory access and reduce trainable parameters without compromising the generalizability.

TABLE VIII
LPDDR VS HBM PERFORMANCE ANALYSIS

| Specification Domain | LPDDR5X | HBM2E |
|---|---|---|
| **Chronological Era** | 2021 (JESD209-5C) | 2021 (JESD235D) |
| **Clock Frequency** (Data Strobe) | 4.26 GHz - 4.8 GHz | 1.6 GHz - 1.8 GHz |
| **Interface / Bus Width** | 64-bit per package | 1024-bit per stack |
| **Max Bandwidth** (Single Package) | 68.2GB/s - 76.8GB/s | 410GB/s - 460 GB/s |
| **Core Voltage (VDD)** | 1.05V | 1.2V |
| **Energy per bit (pJ/bit)** | ~2.0 pJ/bit to 4.1 pJ/bit | ~3.5 pJ/bit to 4.5 pJ/bit |

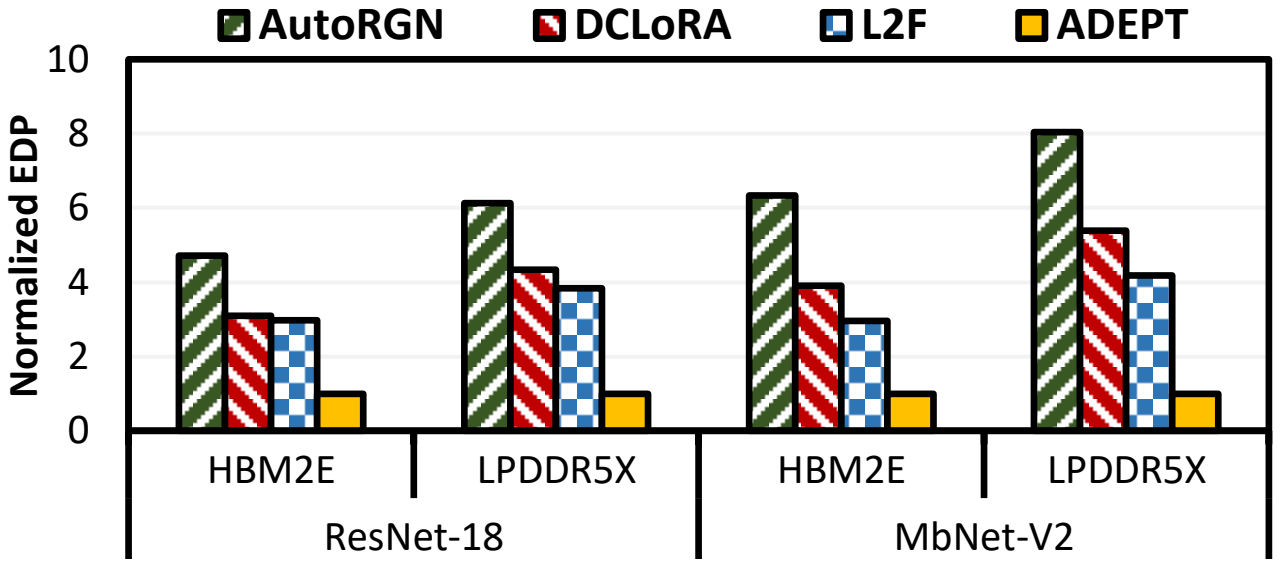

**Fig. 11:** Comparative analysis of EDP gain in ResNet-18 and MbNet-V2 considering CIFAR10 fine-tuning dataset with varying DRAM technology.

**DRAM Technology Sensitivity:** The preceding evaluation utilizes HBM2E to ensure consistency with existing PIM baselines [13], [21], [38]. We now evaluate ADEPT considering different DRAM technologies and explicitly demonstrate that ADEPT's benefits are not specific to a memory technology. We compare the EDP using both LPDDR5X [43] and HBM2E parameters [41] (see Table VIII), keeping all other platform parameters fixed. Fig. 11 reports the resulting normalized EDP for ResNet-18 and MbNET-V2 on PUMA considering CIFAR10 dataset as an example. ADEPT's relative advantage is preserved under LPDDR5X, since off-chip data transfer dominates the end-to-end fine-tuning latency and energy considering any DRAM technology for fine-tuning on PIM architectures. In fact, LPDDR5X's lower bandwidth results in a slightly higher normalized saving compared to HBM2E (7.1× under HBM2E to 8.4× under LPDDR5X on ResNet-18). Overall, ADEPT's gains are derived from reducing the volume of off-chip traffic (see Fig. 9(a)), a mechanism independent of the underlying DRAM technology. Additionally, this finding is valid for any future generation of DRAM technologies too (both LPDDR and HBM).

### *F. Robustness to Continual Learning*

Finally, we evaluate the online sensitivity adaptation introduced in Section IV.E under a continual learning scenario, where the input distribution is non-stationary. We consider ResNet-18, pre-trained on CIFAR10, and expose the model to a stream of CIFAR10-c corruptions encountered sequentially (frost → snow → fog), a common benchmark in the continual learning literature [23]. In our setup, we carry the adapted checkpoint from one stage into the next, so that each corruption constitutes a distribution shift relative to the model's current state. Each task is fine-tuned using the hyperparameters in Table I, utilizing an EMA smoothing factor of $\lambda = 0.5$, with the SER ranking re-evaluated natively at epoch boundaries. We use k = 1 ($\lambda = 0.5$), which weights the new sample and the running estimate equally and implements the scaling as a single right-shift without any multiplier operation.

To highlight the necessity of dynamic tracking, we evaluate three distinct adaptation policies. AutoRGN acts as our upper-bound predictive accuracy reference, adjusting layer-wise learning rates dynamically but necessitating the simultaneous updating of all blocks at full-network EDP cost. ADEPT-static represents a naive hardware-aware baseline, which calibrates the SER ranking only once during the initial corruption and strictly retains it. Finally, ADEPT harnesses the on-chip EMA to refresh block sensitivity online, stochastically re-prioritizing block as the distribution shifts without any host CPU intervention or explicit task-boundary detection.

Table IX reports, for each corruption stage, the accuracy before (No FT) and after the fine tuning (FT). Note that all fine-tuning techniques begin from the identical checkpoints. Given our setup, ADEPT-static and ADEPT yield identical accuracy on the initial "frost" corruption by design. However, the static prioritization grows stale on the subsequent corruptions and degrades because the priorities calibrated on frost no longer reflect the shifted sensitivity profile. Thus, the blocks required to capture the "snow" and "fog" data distributions are stochastically frozen too frequently, resulting in a reduced accuracy recovery of up to 3% relative to the AutoRGN baseline. In contrast, ADEPT successfully refreshes the SER metric using the available gradients. It dynamically unfreezes the most critical layers, tracking AutoRGN to within a worst-case margin of 0.3% across every stage. Crucially, it matches the train-all-blocks accuracy of AutoRGN at a small fraction of its fine-tuning cost. Overall, ADEPT sustains the hardware-aware block selection under evolving data distributions while requiring no host CPU intervention and remains applicable under gradual data distribution drift.

TABLE IX
ACCURACY (%) UNDER CONTINUAL LEARNING SCENARIO

| | FROST | | SNOW | | FOG | |
|---|---|---|---|---|---|---|
| | No FT | FT | No FT | FT | No FT | FT |
| AutoRGN | 60% | 81.1% | 75.2% | 87% | 77.4% | 83.8% |
| ADEPT-Static | 60% | 80.8% | 75% | 83.9% | 67% | 80.2% |
| ADEPT | 60% | 80.8% | 75% | 86.8% | 74.9% | 83.9% |

Finally, we quantify the EDP overhead of this online tracking mechanism. The per-block EMA update uses operands that are already streaming through the DRAM for weight updates and circumvents any additional off-chip DRAM fetches. The subsequent SER update is performed at the pre-existing pipeline-drain boundary at the end of an epoch. Given the baseline fine-tuning EDP on PIM architectures is dominated by off-chip DRAM accesses [13], [44], the localized scalar arithmetic for the EMA leaves the energy term essentially unchanged. Consequently, the tracking overhead is purely bound by a negligible latency increase, translating to under 2% additional end-to-end EDP relative to ADEPT-static.

## VI. CONCLUSION

This paper presented ADEPT, a fine-tuning framework for PIM-based accelerators. The hardware-aware selection, combined with a dynamic channel-wise LoRA in ADEPT, enables energy-efficient fine-tuning. We demonstrate that hardware-aware fine-tuning can significantly outperform sensitivity-only strategies in terms of performance and energy-efficiency, indicating that on-device learning must co-consider model and hardware characteristics. Specifically, the proposed SER (Sensitivity-EDP Ratio) metric produces hardware architecture-dependent block prioritizations that differ substantially between memory-centric PIM and compute-centric platforms, confirming that PIM's asymmetric on-chip/off-chip cost structure necessitates hardware-aware fine-tuning strategies. ADEPT further sustains its hardware-aware selection under non-stationary, continual-learning deployments through an online sensitivity update. Our evaluations across multiple PIM-based architectures show ADEPT achieves up to 8.1× EDP reduction over full-parameter training.

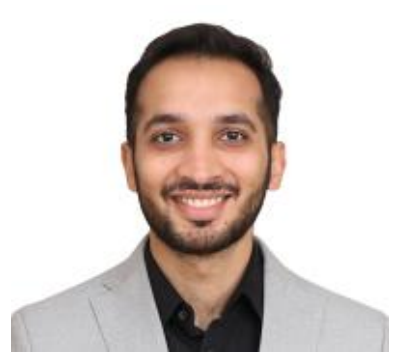

**Pratyush Dhingra** (Graduate Student Member, IEEE) is currently pursuing his Ph.D. degree in Computer Engineering at Washington State University, Pullman, WA, USA. His research interests include memory-centric architectures, AI/HPC accelerators and resource-aware ML.

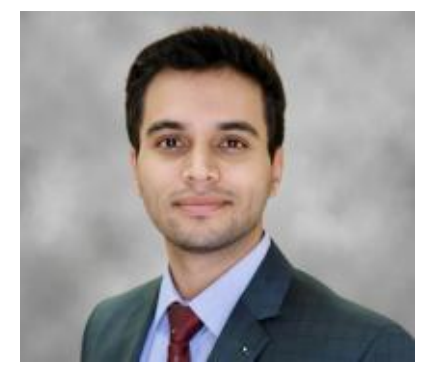

**Vibhanshu Sharma** (Graduate Student Member, IEEE) is currently pursuing the Ph.D. degree in Computer Engineering at Washington State University, Pullman, WA, USA. His research interests include Dynamic Resource Management, ML for EDA, and reliable in-memory processing architectures.

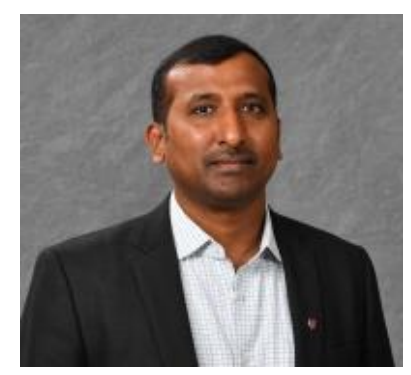

**Janardhan Rao Doppa** (Senior Member, IEEE) received the Ph.D. degree in computer science from Oregon State University, Corvallis, OR, USA, in 2014.
He is the Huie-Rogers Endowed Chair Professor of Computer Science and Berry distinguished Professor in Engineering at Washington State University, Pullman, WA, USA.

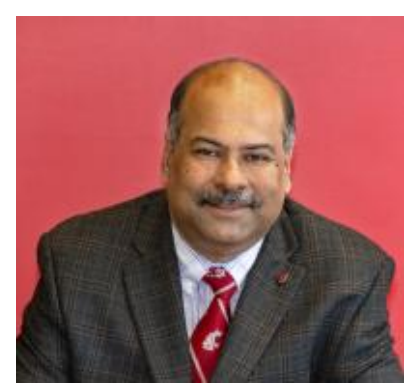

**Partha Pratim Pande** (Fellow, IEEE) received the Ph.D. degree in electrical and computer engineering from the University of British Columbia, Vancouver, BC, Canada, in 2005.
He is a Professor and Dean of Voiland College Of Engineering and Architecture at Washington State University, Pullman, WA, USA.